\documentclass[twocolumn,astrosymb]{aastex631}

\usepackage{amsmath,amssymb}

\newcommand{\bmx}{\boldsymbol{x}}
\newcommand{\bmk}{\boldsymbol{k}}

\newcommand{\bea}{\begin{eqnarray}}
\newcommand{\eea}{\end{eqnarray}}
\newcommand{\dgwa}{\delta^{w_1}}
\newcommand{\dgwb}{\delta^{w_2}}
\newcommand{\dgwc}{\delta^{w_3}}

\newcommand{\epl}{\epsilon}

\newcommand{\bma}[1]{\boldsymbol{[1]}}
\newcommand{\Mpc}{\mathrm{Mpc}}

\shorttitle{Halo tides}
\shortauthors{Zhu et al.}
\graphicspath{{./}{figures/}}

\begin{document}

\title{Cosmic Tidal Reconstruction with Halo Fields}


\author[0000-0002-8202-8642]{Hong-Ming Zhu}
\email{hmzhu@cita.utoronto.ca}
\affiliation{
Canadian Institute for Theoretical Astrophysics, University of Toronto, 60 St. George Street, Toronto, Ontario M5S 3H8, Canada
}

\author[0000-0001-6772-9814]{Tian-Xiang Mao}
\affiliation{National Astronomical Observatories, Chinese Academy of Sciences,
20A Datun Road, Beijing 100101, China}

\author{Ue-Li Pen}%
\affiliation{
Canadian Institute for Theoretical Astrophysics, University of Toronto, 60 St. George Street, Toronto, Ontario M5S 3H8, Canada
}
\affiliation{%
 Dunlap Institute for Astronomy and Astrophysics, University of Toronto, 50 St. George Street, Toronto, Ontario M5S 3H4, Canada
}%
\affiliation{%
 Canadian Institute for Advanced Research, CIFAR Program in Gravitation and Cosmology, \\
 Toronto, Ontario M5G 1M1, Canada
}%
\affiliation{%
 Perimeter Institute for Theoretical Physics, 31 Caroline Street North, Waterloo, Ontario N2L 2Y5, Canada
}%

\begin{abstract}
The gravitational coupling between large-scale perturbations and small-scale perturbations leads to anisotropic distortions of the small-scale matter distribution.
The measured local small-scale power spectrum can thus be used to infer the large-scale matter distribution. 
In this paper, we present a new tidal reconstruction algorithm for reconstructing large-scale modes using the full three-dimensional tidal shear information.
We apply it to simulated dark matter halo fields and the reconstructed large-scale density field correlates well with the original matter density field on large scales, improving upon the previous tidal reconstruction method which only uses two transverse shear fields.
This has profound implications for recovering lost 21~cm radial modes due to foreground subtraction and constraining primordial non-Gaussianity using the multi-tracer method with future cosmological surveys.
\end{abstract}



\section{Introduction} \label{sec:intro}

Galaxy redshift surveys have been a powerful probe of cosmology and fundamental physics.
The measurement of galaxy power spectrum and correlation function has placed stringent constraints on the dark energy properties, modified gravity, primordial non-Gaussianity, etc \citep[e.g.][]{2017MNRAS.470.2617A,2019JCAP...09..010C,2021PhRvD.103h3533A,2021arXiv210613725M}.
The galaxy spectroscopic redshift surveys in the near future will continue to improve the volume and the number of observed galaxies (e.g. DESI \citet{2016arXiv161100036D}; Euclid \citet{2018LRR....21....2A,2020A&A...642A.191E}; SPHEREx \citet{2014arXiv1412.4872D}; MegaMapper \citet{2019BAAS...51g.229S,2019BAAS...51c..72F}).
These experiments will have much lower shot noise level and are able to observe structures on much smaller scales.
However, the strong non-Gaussian nature of the small scale densities makes it challenging to extract cosmological information from observations.
To obtain better constraints on cosmological parameters with future surveys, it is useful to develop new methods to better exploit the higher order statistics.

The matter distribution in the Universe show striking non-Gaussian features such as filaments, voids, superclusters, etc, as a result of nonlinear structure formation.
The initial nearly Gaussian perturbations on different scales couple together due to gravitational instabilities.
The gravitational interactions of the long wavelength density perturbation with small-scale perturbations lead to distortions of the locally measured power spectrum, depending on the local values of the large-scale tidal field.
The trace part of the tidal field describes the change of mean local density variance, leading to the change of small-scale matter power spectrum amplitude, while the trace free part describes the residual anisotropy of gravitational forces, leading to anisotropic distortions of small-scale density perturbations.
Since the local small-scale matter power spectrum is statistically isotropic in the absence of the mode coupling, these local anisotropic tidal distortions can be used to infer the large-scale density distributions.
The reconstruction of gravitational tidal fields from small-scale density field is described by the same formulation as the reconstruction of gravitational lensing from the cosmic microwave background 
\citep{1999LensingEst,1999LensingEst2,2012LensingEst,2018LensingEst,2020arXiv201108251Z} and 21~cm intensities \citep{2008LensingEst,2010LensingEst}.

The purely transverse shear tidal reconstruction has been first presented in \citet{2012Tides} and further investigated by \citet{2016Tides}.
The estimated large-scale density field is given by a linear convolution of the reconstructed tidal shear fields \citep{2012Tides,2016Tides}.
The estimator can also be constructed using the nonlinear coupling coefficient from standard perturbation theory \citep{2018JCAP...07..046F,2020PhRvD.101h3510L,2020arXiv200700226L,2020arXiv200708472D}.

In 21~cm intensity mapping experiments, the radial modes on large scales can not be measured due to the continuum foreground contamination.
This method enables recovery of lost radial modes in 21~cm intensity 
mapping surveys \citep{2018PhRvD..98d3511Z,2018JCAP...07..046F,2019PhRvD.100b3517L,2019MNRAS.486.3864K}, which is important for cross correlating 21~cm intensity fields with weak gravitational lensing, photometric galaxies, kinematic Sunyaev-Zel'dovich effect, etc \citep[e.g.][]{2018PhRvD..98d3511Z,2019PhRvD.100b3517L}.
The reconstructed field provides an independent tracer of the large-scale matter density field.
Thus, it also enables the use of sample variance cancellation technique to better measure cosmological parameters \citep[][]{2009MT,2009MT2}, e.g., primordial non-Gaussianity \citep{2020arXiv200708472D}.

In this paper, we present a new algorithm for tidal reconstruction.
The purely transverse tidal reconstruction uses only two transverse shear terms to reconstruct the large-scale density field \citep[see e.g.][]{2012Tides,2016Tides,2019MNRAS.486.3864K}.
Here we construct estimators for another three tidal shear fields.
We apply the new method to dark matter halos from high precision simulations and find the reconstructed field is well correlated with the large-scale matter density field on large scales.
The tidal reconstruction method is thus a proven useful way for extracting information from small-scale density fluctuations and helpful for recovering lost 21~cm radial modes and multi-tracer analysis.

This paper is organized as follows.
In Section \ref{sec:formalism}, we describe the formalism of tidal reconstruction.
In Section \ref{sec:implemention}, we present the numerical implementation of tidal reconstruction.
In Section \ref{sec:results}, we apply the method to halos samples from simulations and show the numerical results. 
We discuss and conclude in Section~\ref{sec:dissconcl}.

\section{Formalism}
\label{sec:formalism}

The gravitational interaction between the long wavelength perturbation and small-scale
density fluctuations has been studied generally \citep[e.g.][]{2014Tides,2016Tides}.
In this section, we give a brief description of the effect arising from tidal interaction, and present the formalism for tidal reconstruction.
A numerical implementation of the reconstruction algorithm will be presented in the next section.

We consider the gravitational mode coupling of a long wavelength perturbation with small-scale density fluctuations in the squeezed limit, i.e., the wavelength of small-scale density perturbations is sufficiently smaller than that of the large-scale density perturbation.
The leading order observable is then described by the large-scale tidal field,
\begin{equation}
t_{ij}=\Phi_{L,ij},
\end{equation}
where $\Phi_L$ is the long wavelength gravitational potential sourced by the
long wavelength density perturbation $\delta_L$.
Here, $\Phi_{L,ij}$ denotes partial derivatives of $\Phi_L$ to $x^i$ and $x^j$.
The effects of the large-scale tidal field on small-scale density 
perturbations can be calculated using perturbation theory.
In the perturbative calculation, the gravitational potential which drives the
motion of a particle is sourced by both the small-scale density fluctuations 
and the tidal field,
\begin{equation}
\phi(\bmx,\tau)=\Phi_s(\bmx,\tau)+\frac{1}{2}t_{ij}(\boldsymbol{0},\tau)x^ix^j,
\end{equation}
and
\begin{equation}
t_{ij}(\boldsymbol{0},\tau)=T(\tau)t_{ij}^{(0)}(\boldsymbol{0}),
\end{equation}
where $\boldsymbol{x}$ is the comoving Eulerian coordinate, $\tau$ is the conformal time,
$T(\tau)=D(\tau)/a(\tau)$ is the linear transfer function, $D(\tau)$ is the 
linear growth function, $a(\tau)$ is the scale factor, the superscript $(0)$ 
denotes the tidal field evaluated at some initial time $\tau_0$, and 
$D(\tau_0)=a(\tau_0)=1$.

Using the second-order perturbation theory, the leading order contribution to
the density contrast from the tidal field is given by
\begin{eqnarray}
\label{eq:deltat}
\delta_t(\bmx,\tau)=t^{(0)}_{ij}\bigg[\alpha(\tau)\frac{\partial^i\partial^j}
{\nabla^2}+\beta(\tau)x^i\partial^j+\gamma(\tau)\delta^{ij}\bigg]
\delta_{1s}(\bmx,\tau), \nonumber \\
\end{eqnarray}
where $\alpha(\tau)$, $\beta(\tau)$, and $\gamma(\tau)$ are tidal coupling 
coefficients which describe the strength of the tidal interaction. 
These coefficients depend on the background cosmology and can be computed 
analytically \citep[see][for details]{2014Tides,2016Tides}.
The density field $\delta_{1s}$ is the linear density contrast computed
using linear theory without the external tidal field $t_{ij}$.
In the Einstein-de Sitter Universe, the above expression corresponds to the standard second-order perturbation 
result in the squeezed limit that the wavelength of one mode is much larger than that 
of the other one \citep{2014Tides}.

Let us consider a small patch of the Universe with scale much smaller than the 
wavelength of the tidal field. 
Then the tidal field $t_{ij}$ can be considered as spatially constant in this
small region. 
The small-scale matter correlation function is given by
\bea
\xi(\boldsymbol{r},\tau)=\langle\delta(\boldsymbol{0},\tau)\delta(\boldsymbol{r},\tau)\rangle,
\eea
where 
\bea
\delta(\bmx,\tau)=\delta_{1s}(\bmx,\tau)+\delta_{t}(\bmx,\tau).
\eea
Inserting Eq. (\ref{eq:deltat}), we have
\bea
\label{eq:xi}
\xi(\boldsymbol{r},\tau)&=&\xi_{1s}(r,\tau)+t^{(0)}_{ij}\times\nonumber\\
                &&\bigg[2\alpha(\tau)
    \frac{\partial^i\partial^j}{\nabla^2}
    +\beta(\tau)x^i\partial^j
+2\gamma(\tau)\delta^{ij}\bigg]\xi_{1s}(r,\tau),\nonumber\\
\eea
where $\xi_{1s}(r,\tau)$ is the isotropic linear matter correlation function.
From Eq. (\ref{eq:xi}), it is clear that the external tidal field induces the 
anisotropic distortion of the locally measured matter correlation function.
We then transform Eq. (\ref{eq:xi}) to Fourier space to obtain the locally
observed matter power spectrum
\bea
\label{eq:pk}
P(\bmk,\tau)|_{t_{ij}}&=&P_{1s}(k,\tau)\nonumber\\
&&+t^{(0)}_{ij}
\bigg[2\alpha(\tau)\hat{k}^i\hat{k}^j-\beta(\tau)
\frac{d\ln P_{1s}(k,\tau)}{d\ln k}\hat{k}^i\hat{k}^j\nonumber\\
&&+\big(2\gamma(\tau)-\beta(\tau)\big)\delta^{ij}
\bigg]P_{1s}(k,\tau),
\eea
where $P_{1s}(k,\tau)$ is the isotropic linear power spectrum and $\hat{\boldsymbol{k}}$
is the unit vector.

The $3\times3$ symmetric tensor field $t_{ij}$ can be decomposed into six 
orthogonal observables,
\bea
t_{ij}=\left( \begin{array}{ccc}
        \epl_0+\epl_1-\epl_z & \epl_2 & \epl_x \\
        \epl_2 & \epl_0-\epl_1-\epl_z & \epl_y \\  
        \epl_x & \epl_y & \epl_0+2\epl_z
\end{array} \right),
\eea
where $\epl_{0}=(\Phi_{L,11}+\Phi_{L,22}+\Phi_{L,33})/3$, $\epl_{1}=(\Phi_{L,11}-\Phi_{L,22})/2$, $\epl_{2}=\Phi_{L,12}$, $\epl_{x}=\Phi_{L,13}$, $\epl_{y}=\Phi_{L,23}$, and $\epl_{z}=(2\Phi_{L,33}-\Phi_{L,11}-\Phi_{L,22})/6$.
The trace part of the tidal field corresponds to the local mean density, while
the other components describe the residual anisotropic gravity shear forces.

With such decomposition, the local power spectrum can be written as
\bea
P(\bm{k},\tau)&=&P_{1s}(k,\tau)+\Big[
f_0(k,\tau)\epl_0
+f(k,\tau)\Big((\hat{k}_1^2-\hat{k}_2^2)\epl_1\nonumber\\&&
+2\hat{k}_1\hat{k}_2\epl_2
+2\hat{k}_1\hat{k}_3\epl_x
+2\hat{k}_2\hat{k}_3\epl_y
+\big(2\hat{k}_3^2
-\hat{k}_1^2\nonumber\\&&
-\hat{k}_2^2\big)\epl_z\Big)
\Big]P_{1s}(k,\tau).
\eea
where 
\bea
f_0(k,\tau)=f(k,\tau)+6\gamma(\tau)-3\beta(\tau),
\eea
and
\bea
f(k,\tau)=2\alpha(\tau)-\beta(\tau)\frac{d\ln P_{1s}(k,\tau)}
{d\ln k}.
\eea
The coupling coefficient for shear is different from that of the tidal field trace.
Notice that the coupling coefficient is scale-independent for the power-law spectrum $\propto k^n$, as long as the mode coupling is in the squeezed limit.

The tidal field trace which corresponds to the local mean density changes the amplitude of small-scale power spectrum.
The traceless shear components then causes anisotropic distortions to the local small-scale power spectrum.
The anisotropic tidal distortions have different angular dependence.
The two tensor-like tidal shear fields $\epl_1$ and $\epl_2$ are characterized by $\cos2\phi$ and $\sin2\phi$ dependence, respectively, where $\phi$ is the azimuthal angle in Fourier space. 
The two vector-like shear fields $\epl_x$ and $\epl_y$ have angular dependence described by $\sin\theta\cos\phi$ and $\sin\theta\sin\phi$, respectively, where $\theta$ is the polar angle in Fourier space.
The scalar-like tidal shear field $\epl_z$ which describes the distortions along the $z$ axis has $(3\cos^2\theta-1)/2$ dependence which preserves the azimuthal
symmetry about the $z$-axis direction.

The idea of cosmic tidal reconstruction is to measure these local anisotropic tidal distortions by integrating the locally measured small-scale power spectrum with some angular-dependent weights.
The tidal coupling results in a systematic change of the small-scale power.
Since the large scale tidal field is coherent on small scales, when enough small-scale modes are measured, the signal are dominated by the large-scale tidal shear.
The linear convolution of the different tidal shear fields gives the estimate
of the large-scale density field.
This is the same as lensing reconstruction using cosmic microwave background 
\citep{1999LensingEst,1999LensingEst2,2012LensingEst,2018LensingEst,2020arXiv201108251Z} and 21~cm temperature field \citep{2008LensingEst,2010LensingEst}.
In practice, this is accomplished by the use of quadratic estimators, which will be discussed in the next section.

\section{Implementation}
\label{sec:implemention}

In this section, we describe the implementation of tidal reconstruction and discuss the approximations we made in deriving the tidal field estimator.

The large-scale tidal shear fields can be estimated with the quadratic 
estimators, which are outer products of the filtered density fields \citep{2008LensingEst,2010LensingEst,2012LensingEst},
\bea
\label{eq:est}
\hat{\epl}_1(\bmx)&=&\big[\dgwa(\bmx)\dgwa(\bmx)
-\dgwb(\bmx)\dgwb(\bmx)\big]/2,\nonumber\\
\hat{\epl}_2(\bmx)&=&\dgwa(\bmx)\dgwb(\bmx),\nonumber\\
\hat{\epl}_x(\bmx)&=&\dgwa(\bmx)\dgwc(\bmx),\nonumber\\
\hat{\epl}_y(\bmx)&=&\dgwb(\bmx)\dgwc(\bmx),\nonumber\\
\hat{\epl}_z(\bmx)&=&\big[2\dgwc(\bmx)\dgwc(\bmx)-\dgwa(\bmx)\dgwa(\bmx)
\nonumber\\&&
-\dgwb(\bmx)\dgwb(\bmx)\big]/6,
\eea
where
\bea
\label{eq:deltawp}
\delta^{w_j}(\bmk)=i\hat{k}_jw(k)\delta(\bmk),
\eea
is the filtered density field.
In the long wavelength limit and under the Gaussian assumption, the optimal 
window function for minimum variance quadratic estimator can be constructed as
\bea
\label{eq:window}
w(k)=\frac{\sqrt{P(k)f(k)}}{P_{\mathrm{tot}}(k)},
\eea
where $P_{\mathrm{tot}}(k)$ is the total power spectrum measured from 
cosmological observations which includes both the cosmic signal and noise.
This is only valid with data in the linear and mildly nonlinear regime where perturbation theory can give a good description.
However, most observable modes in galaxy surveys are in the nonlinear regime and thus outside the realm of perturbation theory.
To construct estimators strictly following the perturbation theory prediction, these small-scale modes in the nonlinear regime are discarded in reconstruction or smoothed on quasi-linear scales \citep[e.g.][]{2018JCAP...07..046F,2020PhRvD.101h3510L,2020arXiv200700226L,2020arXiv200708472D}, which degrades the reconstruction performance substantially.

However, if we consider only the squeezed limit, the coupling coefficient can be approximated as constant for a power law spectrum $\propto k^n$ with $n\approx-2$, which is an approximation for nonlinear power on small scales.

In this case, Eq. (\ref{eq:deltawp}) becomes
\bea
\label{eq:deltap}
\delta^{w_j}(\bmk)\propto ik_j W_R(k)\delta(\bmk),
\eea
up to a proportional constant, where
\bea
W_R(\bmk)=\exp\left(-k^2R^2/2\right),
\eea
is a Gaussian window function with smoothing scale $R$.
Note that we have replaced the Wiener filter $P(k)/P_\mathrm{tot}(k)$ by the Gaussian window function $W_R(k)$, which reduces the small scale noise.

Since the performance of reconstruction is dominated by small-scale modes on mildly nonlinear scales, this approximated estimator in the long wavelength limit allows us to utilize smaller scales in reconstruction and thus reduce the reconstruction noise.
Although the coupling coefficient or response function can be approximated as constant for the large-scale mode with squeezed momentum and small-scale modes in the nonlinear regime, the constant overall coupling strength is still unknown.
This will lead to a constant bias for the reconstructed field on large scales \citep[see e.g.][]{2016Tides}.
For reconstructed modes with $k\ga0.1\;\mathrm{Mpc}/h$, where the squeezed limit is not satisfied, this will induce some scale-dependent bias, which needs to be corrected with simulations or mock catalogs.

As we discussed in the previous section, all these five shear fields are functions
of the large-scale gravitational potential or the large-scale 
density related through the Poisson equation.
Considering the different angular dependence of the shear fields and using the 
appropriate angular-dependent weight for each shear field, the large-scale
density field is given by the linear convolution of the estimated shear fields,
\bea
\label{eq:deltar_3D}
\epl_0(\bmk)&=&\frac{1}{2k^2}
\Big[(k_1^2-k_2^2)\epl_1(\bmk)+2k_1k_2\epl_2(\bmk)+2k_1k_3\epl_x(\bmk)
\nonumber\\&&
+2k_2k_3\epl_y(\bmk)+(2k_3^2-k_1^2-k_2^2)\epl_z(\bmk)\Big],
\eea
where $\epl_0=\nabla^2\Phi_L/3$ that differs from the density by a constant factor.
In \cite{2012Tides,2016Tides}, we use the two purely transverse shear 
fields $\epl_1$ and $\epl_2$ for density reconstruction in analogy with the weak
lensing mass reconstruction \citep[][]{1993ApJ...404..441K}.
In that case, the large-scale density field is given by
\bea
\epl_0(\bmk)\propto\frac{k^2}{(k_\perp^2)^2}
\Big[(k_1^2-k_2^2)\epl_1(\bmk)+2k_1k_2\epl_2(\bmk)\Big],
\eea
where $k_\perp^2=k_1^2+k_2^2$. 
We note that in general any combination of the shear fields can provide an estimate of the large-scale density field,
\bea
\label{eq:deltar}
\epl_0(\bmk)=\frac{1}{A(\bmk)}\sum_{i\in S}A^i(\bmk)\epl_i(\bmk),
\eea
where $S\subseteq\{1,2,x,y,z\}$ and the angular weights are 
\bea
&A^1(\bmk)=k_1^2-k_2^2,\ A^2(\bmk)=2k_1k_2,\ A^x(\bmk)=2k_1k_3,\nonumber\\&
A^y(\bmk)=2k_2k_3,\ A^z(\bmk)=2k_3^2-k_1^2-k_2^2.\nonumber
\eea
These coefficients weight the anisotropic shear fields appropriately and give 
a proper estimate of the isotropic density field.
The normalization factor is computed as
\bea
A(\bmk)=\frac{3}{k^2}\sum_{i\in S}A^i(\bmk)B_i(\bmk),
\eea
where the coefficients are
\bea
&B_1(\bmk)=(k_1^2-k_2^2)/2,\ B_2(\bmk)=k_1k_2,\ B_x(\bmk)=k_1k_3,\nonumber\\&
B_y(\bmk)=k_2k_3,\ B_z(\bmk)=(2k_3^2-k_1^2-k_2^2)/6,\nonumber
\eea
which are simply pre-factors of the shear fields $\epl_i(\bmk)=-B_i(\bmk)\Phi_L(\bmk)$ in Fourier space. 
In principle, even a single tidal shear field can reconstruct the density field
as long as the estimator is properly normalized.
However, different shear fields are sensitive to different density modes in Fourier space.
Using only two transverse shear fields for tidal reconstruction leads to 
anisotropic reconstruction noise as expected \citep{2012Tides,2016Tides,2019MNRAS.486.3864K}.
We will explore the two different combinations of tidal shears and the resulting anisotropic reconstruction noises in the next section.

In general, the reconstructed density field can be written as
\bea
\delta_r(\bmk)=C(\bmk)\delta(\bmk)+n(\bmk),
\eea
where $\delta_r(\bmk)=\epl_0(\bmk)$ is the density field reconstructed from the shear fields,
$C(\bmk)=P_{\delta_r\delta}(\bmk)/P_\delta(\bmk)$ is the propagator, and 
$\delta(\bmk)$ is the original matter density field.
The propagator quantifies the information of the original density field in the
reconstructed density field. To get an unbiased measurement of the original
density field, we can deconvolve the propagator from the reconstructed field,
\bea
\label{eq:deltar1}
\hat{\delta}_r(\bmk)=\delta_r(\bmk)/C(\bmk)=\delta(\bmk)+N(\bmk),
\eea
where $N(\bmk)=n(\bmk)/C(\bmk)$ is the reconstruction noise.

An alternative method is to compute the transfer function by minimizing the 
difference between the reconstructed and original density fields,
\bea
\langle(t(\bmk)\delta_r(\bmk)-\delta(\bmk))^2\rangle,
\eea
and we have
\bea
\label{eq:tk}
t(\bmk)=\frac{P_{\delta_r\delta}(\bmk)}{P_{\delta_r}(\bmk)}=\frac{1}{C(\bmk)}
\frac{P_\delta(\bmk)}{P_{\hat{\delta}_r}(\bmk)}.
\eea
Applying the transfer function $t(\bmk)$ to the reconstructed density field 
$\delta_r(\bmk)$ is equivalent to applying the Wiener filter $P_{\delta}(\bmk)/P_{\hat{\delta}_r}(\bmk)$ to the unbiased reconstructed density field in Eq. (\ref{eq:deltar1}).
In the case where we include all five shear fields in the reconstruction and without 
redshift space distortions, the propagator and noise only depend on the magnitude of the wave vector $k$.

We have presented the algorithm for tidal reconstruction with tidal shears.
The trace of tidal field $\epl_0(\bmx)$ can be estimated directly using the quadratic estimator in a similar form $\sim\delta^{w_0}_g\delta^{w_0}_g$, where the filter is isotropic such as Gaussian window.
The tidal field trace, i.e., the local mean density, can also be estimated by measuring the amplitude of small-scale power spectrum in different subvolumes \citep[e.g.][]{2014PhRvD..90j3530L,2014JCAP...05..048C,2015JCAP...09..028C}.
The information of tidal shear is orthogonal to the amplitude of local power spectrum, since it comes from the anisotropic distortions of power spectrum.

\section{Results}
\label{sec:results}

To test the reconstruction method, we apply the algorithm to simulations.
The simulations evolve $1536^3$ dark matter particles in a box of size $500\,\mathrm{Mpc}/h$ to redshift $z=0.6$ using the code {\tt MP-Gadget} \citep[][]{yu_feng_2018_1451799}.
The dark matter halos are identified using the standard friends-of-friends algorithm with linking length 0.2 using {\tt nbodykit} \citep[][]{2018AJ....156..160H}. 
Halos are required to include at least 25 dark matter particles, a minimum halo mass of $7.4\times10^{10}\;h^{-1}{M}_\sun$;
the heaviest halo weighs about $1.3\times10^{15}\;h^{-1}{M}_\sun$.
We use three halo catalogs with different minimum mass $\log{M}_\mathrm{min}=10.8,11.8,$ and $12.8\;h^{-1}{M}_\sun$.
The corresponding halo number densities are $4.9\times10^{-2},6.3\times10^{-3},$ and $5.9\times10^{-4}\;(\mathrm{Mpc}/h)^{-3}$.

We use five $N$-body simulations with independent realizations of the linear density.
For each mass bin, we compute the halo densities on a $512^3$ grid using standard cloud-in-cell painting.
We smooth the halo density field with the Gaussian window and then compute the filtered density field using Eq.~(\ref{eq:deltap}).
The smoothing scales are different for the three halo catalogs.
We test a few different scales for each mass bin and find the optimal scales for three number densities for three halo samples are $0.75,1.0,1.25\;\mathrm{Mpc}/h$, with increasing minimum halo mass.
Following Eq.~(\ref{eq:est}), we compute the tidal shear fields.
Finally we obtain the reconstructed field using Eq.~(\ref{eq:deltar_3D}).

\begin{figure*}[htb!]
\centering
\includegraphics[width=0.49\textwidth]{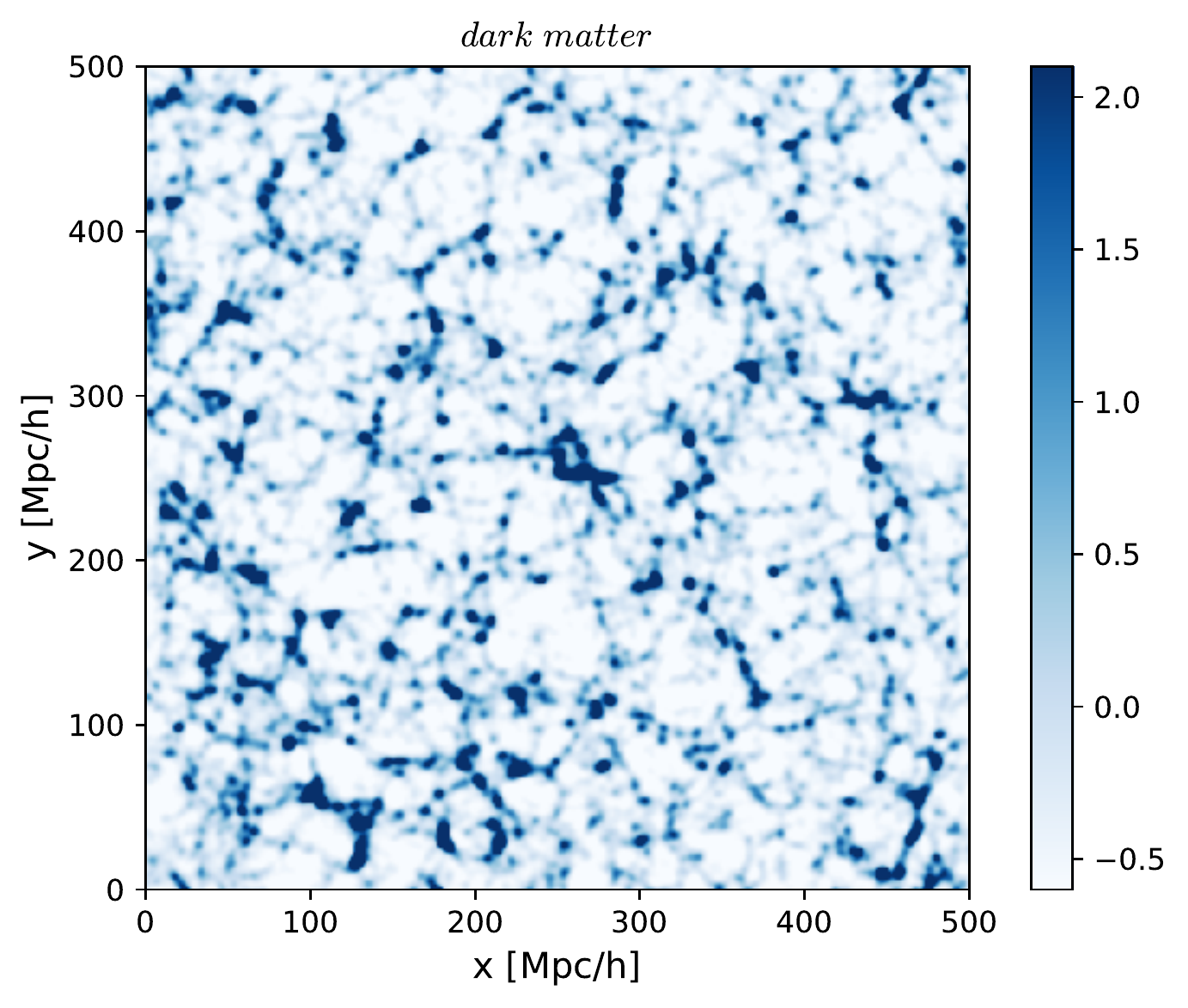}
\includegraphics[width=0.49\textwidth]{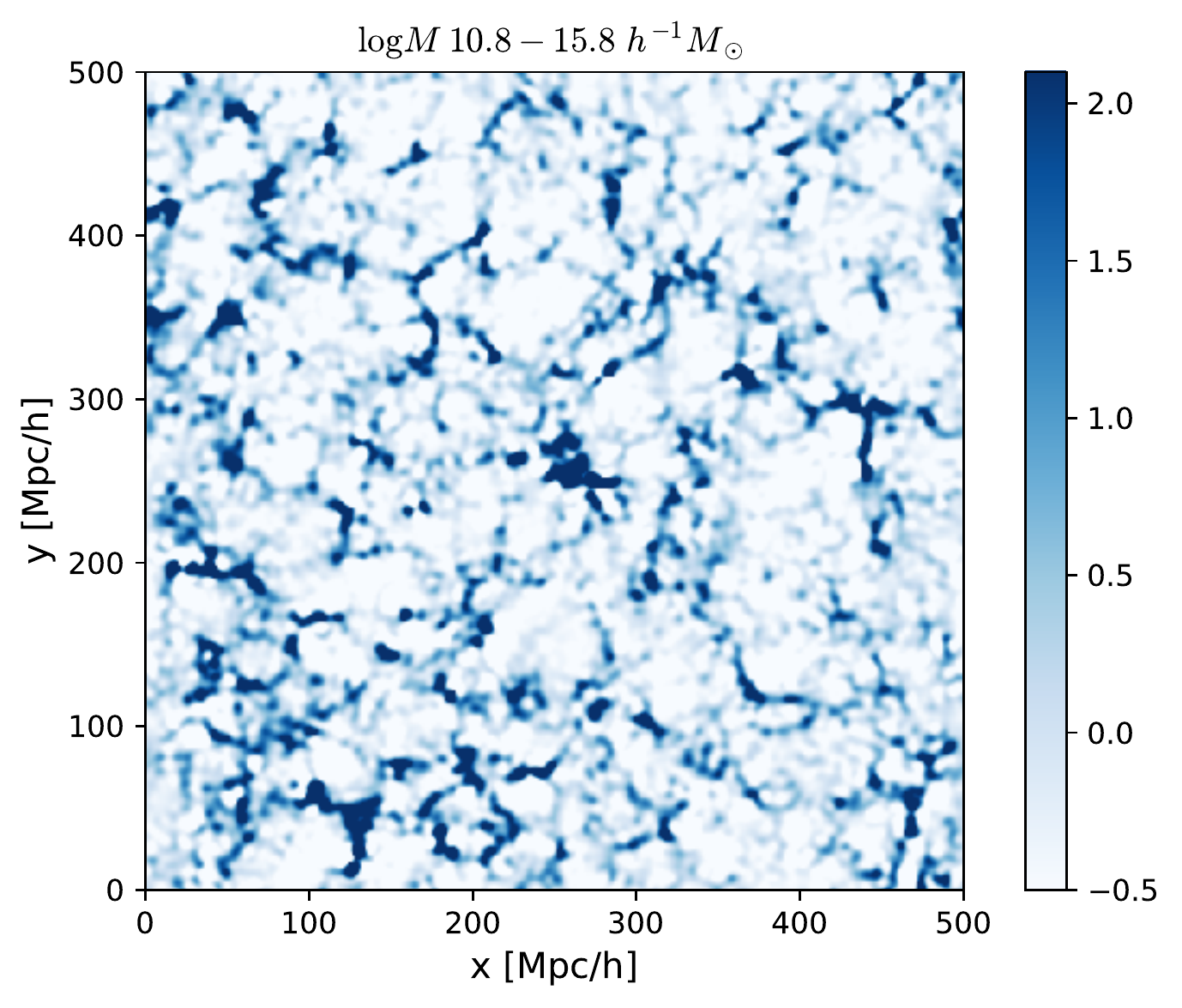}
\includegraphics[width=0.49\textwidth]{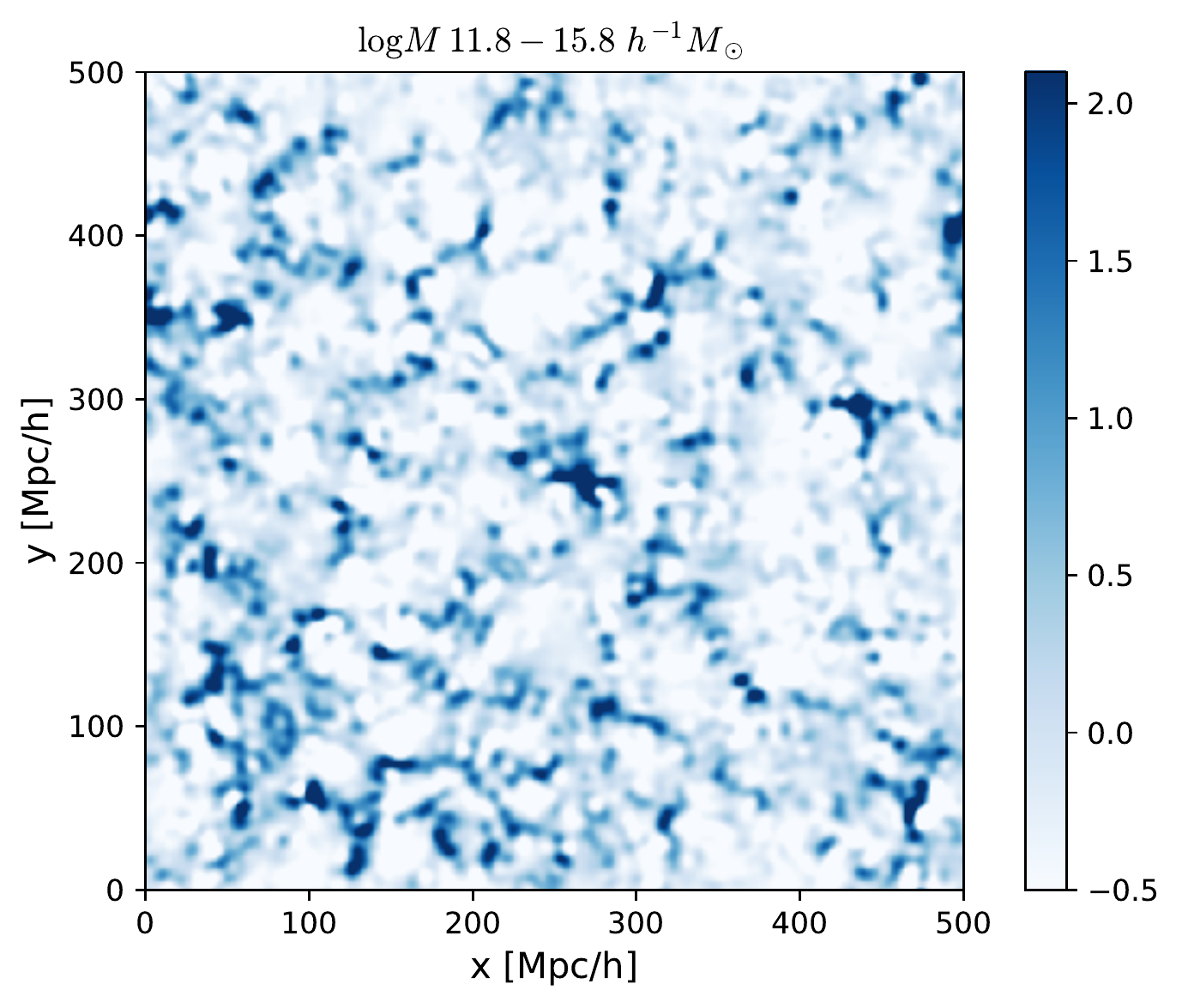}
\includegraphics[width=0.49\textwidth]{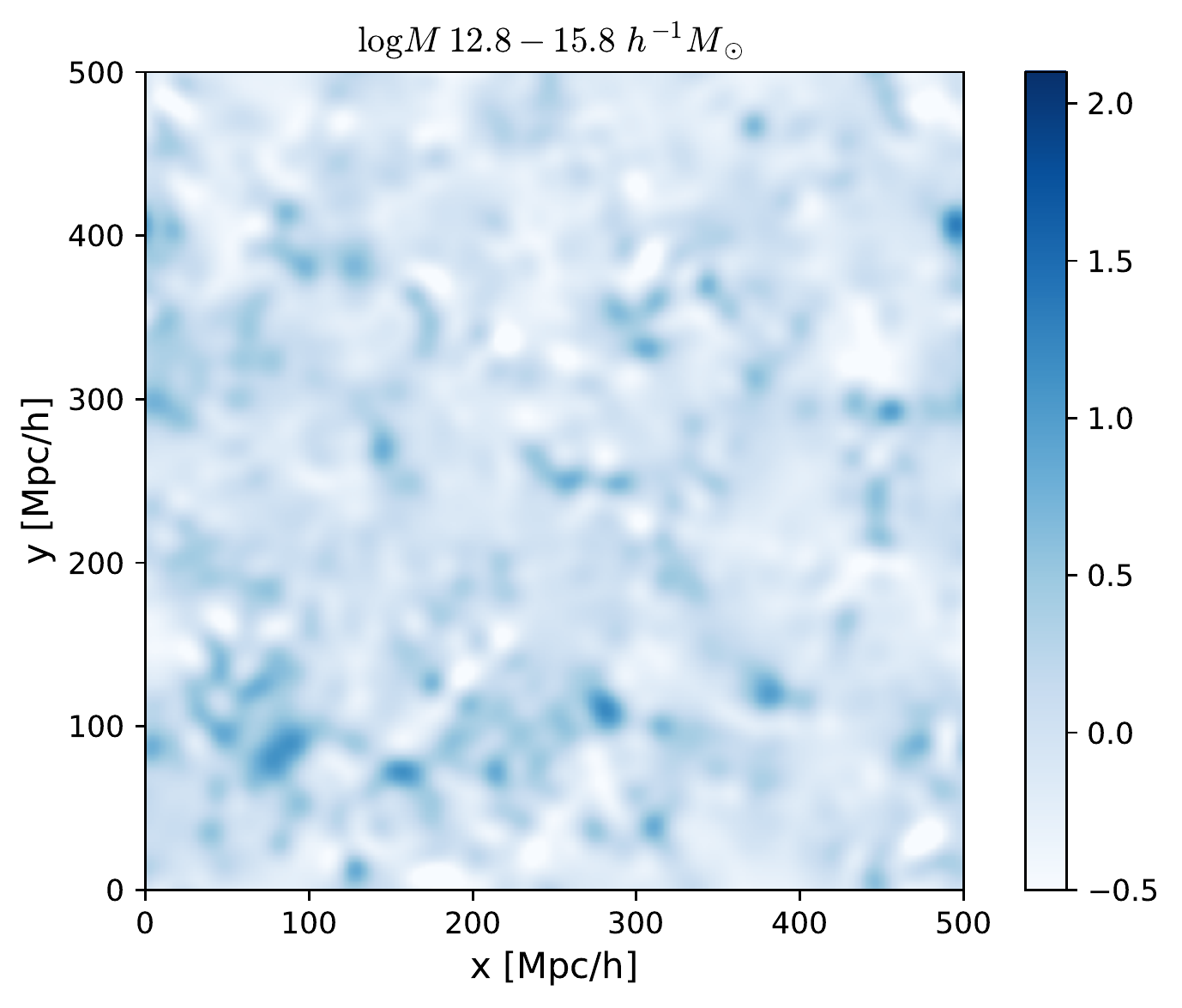}
\caption{
Two-dimensional maps of the dark matter density (Top left), compared with the field reconstructed from halo samples with number density $\bar{n}=4.9\times10^{-2}$ (Top right), $6.3\times10^{-3}$ (Bottom left), and $5.9\times10^{-4}\ (\mathrm{Mpc}/h)^{-3}$ (Bottom right).
The dark matter density is smoothed with a $R=2\;\mathrm{Mpc}/h$ Gaussian, $W_R(k)=\exp(-k^2R^2/2)$.
The reconstructed field is convolved with the transfer function in Eq.~(\ref{eq:tk}) to suppress the small scale noise.
For two higher number density samples (Top right and bottom left), the structure can be reconstructed accurately even on small scales.
For the low number density case (Bottom right), the reconstruction is less accurate, especially on small scales, but still gets most of the large-scale structure right. 
\label{fig:2dmap}}
\end{figure*}

In Fig.~\ref{fig:2dmap},
we plot two-dimensional maps of the dark matter density field and the reconstructed field using tidal reconstruction for three halo samples with minimum mass $\log{M}_\mathrm{min}=10.8,11.8,$ and $12.8\;h^{-1}{M}_\sun$.
The dark matter density field is smoothed on scale $2\;\mathrm{Mpc}/h$ using a Gaussian smoothing.
The reconstructed field is convolved with the transfer function from Eq.~(\ref{eq:tk}) to suppress the small-scale noise.
We see that the maps reconstructed with two higher number density samples show similar structures as the original dark matter density.
The small-scale structure can be reconstructed with high fidelity for higher number density samples. 
This shows that tidal reconstruction provides an accurate estimate of the density contrast of dark matter particles. 
For the lower number density $\sim10^{-4}\;h^3\mathrm{Mpc}^{-3}$, the tidal reconstruction is less successful, providing a less accurate estimate especially on small scales, but still gets most structures on large scales right.

\begin{figure}[htb!]
\includegraphics[width=0.47\textwidth]{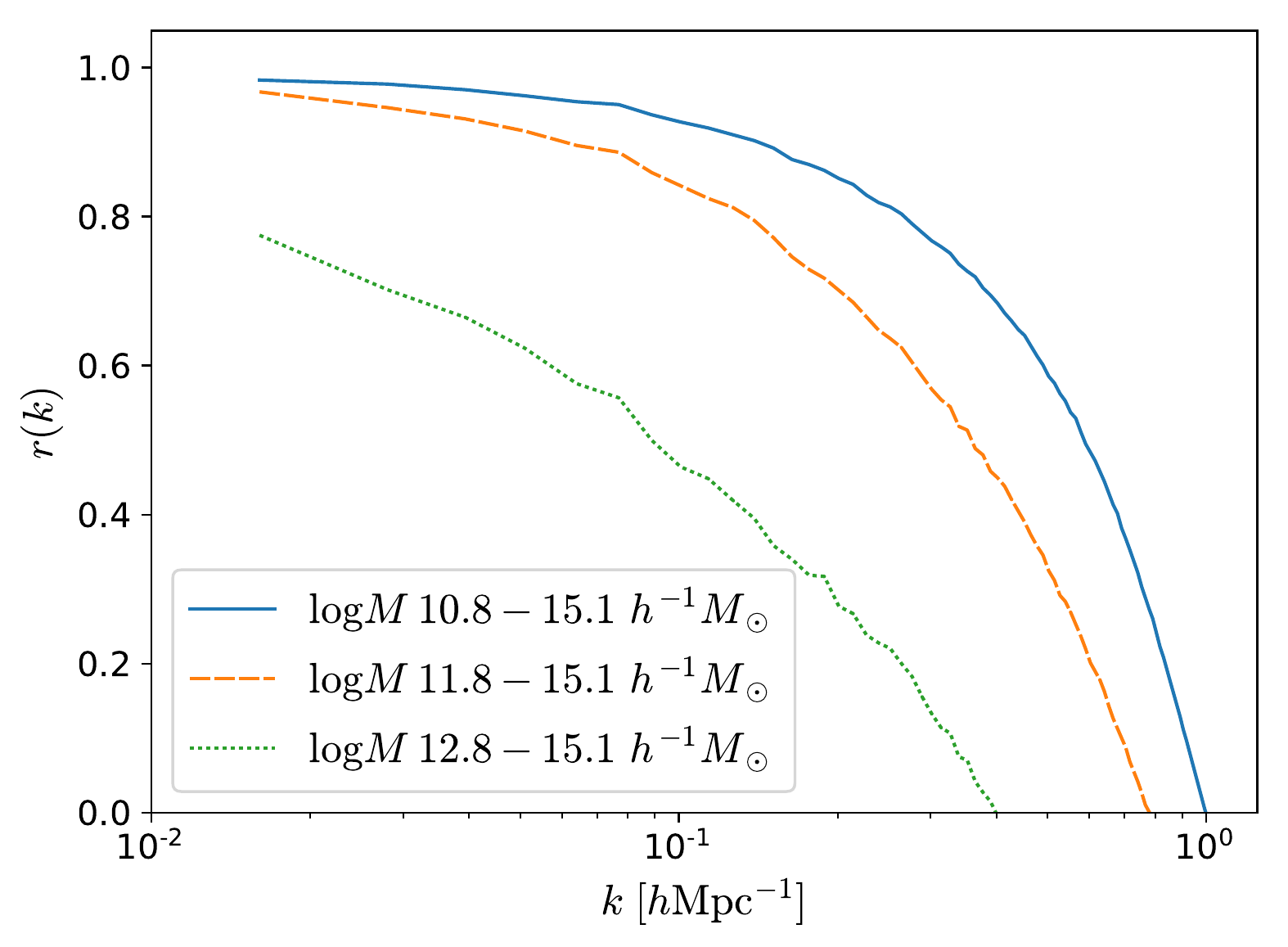}
\caption{
The cross-correlation coefficient $r(k)$ of the reconstructed field $\delta_r$ and the dark matter density field $\delta$ for halo samples with number density $\bar{n}=4.9\times10^{-2}$, $6.3\times10^{-3}$, and $5.9\times10^{-4}\ (\mathrm{Mpc}/h)^{-3}$.
\label{fig:xcc}}
\end{figure}

Figure~\ref{fig:xcc} shows the cross-correlation coefficient of the reconstructed field with the dark matter density field,
\begin{equation}
    r(k)=\frac{P_{\delta_r\delta}(k)}{\sqrt{P_{\delta_r\delta_r}(k)P_{\delta\delta}(k)}},
\end{equation}
where $P_{\delta_r\delta}(k)$ is the cross power spectrum and $P_{\delta_r\delta_r}(k)$ and $P_{\delta\delta}(k)$ are the power spectra of reconstructed field $\delta_r$ and dark matter density $\delta$.
The power spectra are averaged over five simulations first before computing the correlation coefficient.
We find that the cross correlation coefficient is above 0.9 for $\bar n=4.9\times10^{-2}\;h^3\Mpc^{-3}$ and 0.8 for $\bar n=6.3\times10^{-3}\;h^3\Mpc^{-3}$ at $k<0.1\; h\Mpc^{-1}$ and even higher on larger scales.
This shows that the dark matter distribution can be reconstructed with high signal to noise ratio with tidal reconstruction.
While for the lower number density, the correlation coefficient is above $0.5$ at $k<0.1\;h\Mpc^{-1}$ and $80\%$ correlated to the dark matter density field on largest scales, indicating a larger reconstruction noise.
These are in agreement with the intuitive comparison presented in Fig.~\ref{fig:2dmap}.

\begin{figure}[htb!]
\includegraphics[width=0.47\textwidth]{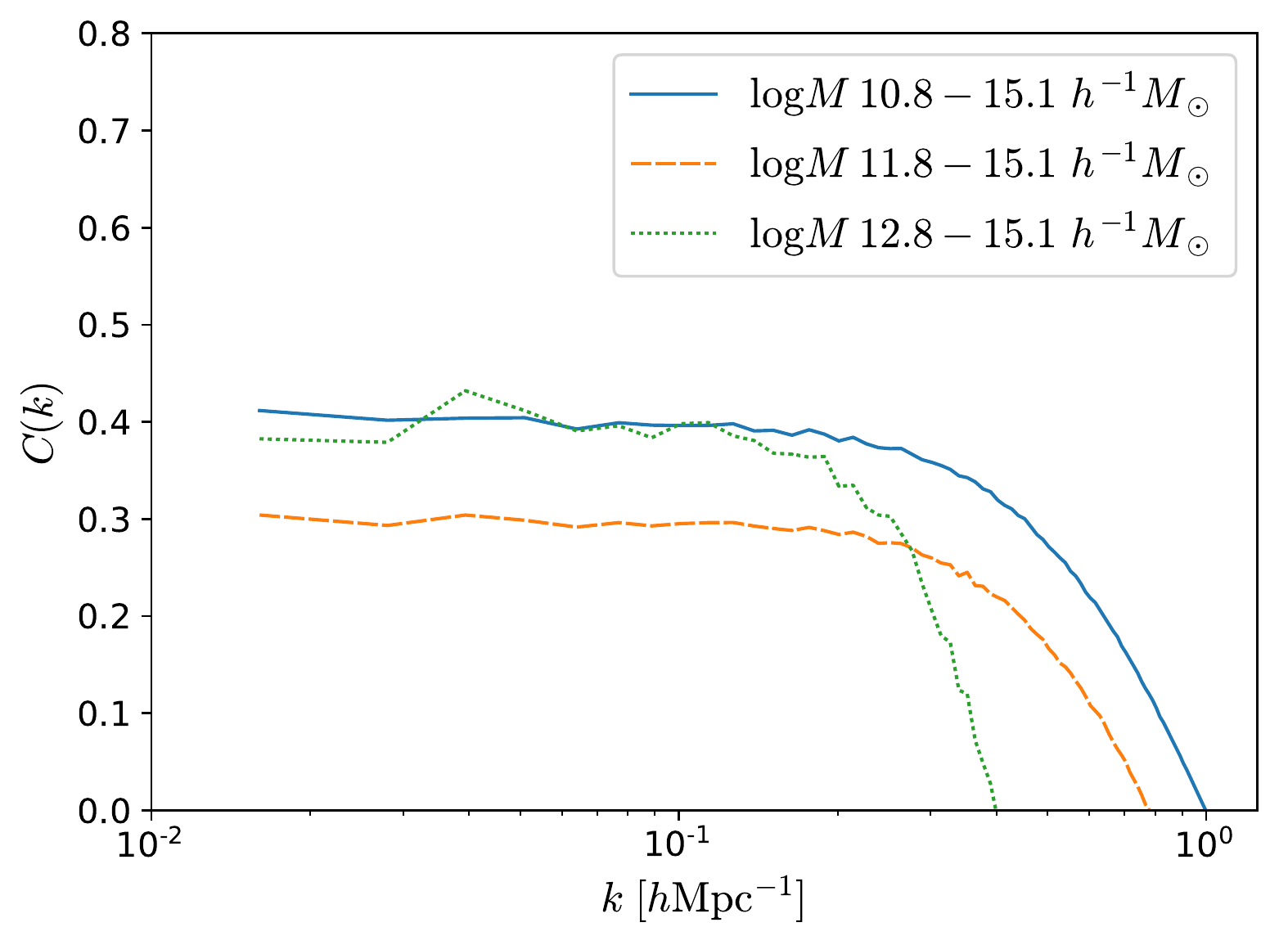}
\caption{The propagator or bias $C(k)$ of the reconstructed field and the dark matter density field for halo samples with number density $\bar{n}=4.9\times10^{-2}$, $6.3\times10^{-3}$, and $5.9\times10^{-4}\ (\mathrm{Mpc}/h)^{-3}$.
\label{fig:br}}
\end{figure}

We have defined the propagator as
\begin{equation}
    C(k)=\frac{P_{\delta_r\delta}(k)}{P_{\delta\delta}(k)},
\end{equation}
which quantifies the bias of the reconstructed field to the original dark matter density field.
Figure~\ref{fig:br} shows the propagator $C(k)$ of the reconstructed fields for three halo mass bins.
The propagator is nearly constant on large scales $k<0.1\;h\mathrm{Mpc}^{-1}$ for the two higher number density halo samples.
We notice that the fluctuations of $C(k)$ exist on large scales for number density $\bar n=5.9\times10^{-4}\;h^3\Mpc^{-3}$, which is likely due to the large reconstruction noise that we will describe shortly.
When the wavenumber $k$ increases, the propagator $C(k)$ deviates from the constant value on large scales and becomes zero at high $k$.
This large-scale constant bias can not be predicted since the constant overall tidal coupling strength is unknown even the tidal response function can be approximated as constant in the squeezed limit.
This bias can be fitted by a linear bias parameter for large-scale modes $k<0.1\;h\Mpc^{-1}$ in the data analysis.
To include higher $k$ modes in the analysis, the scale-dependence of $C(k)$ has to be modeled with mock galaxy catalogs which resembles the clustering properties of specific galaxy samples.
Since the reconstruction is dominated by the large number of small-scale modes, we find that excluding all modes with $k<0.1\;h/\Mpc^{-1}$ in the reconstruction does not affect the results.

\begin{figure}[tb!]
\includegraphics[width=0.47\textwidth]{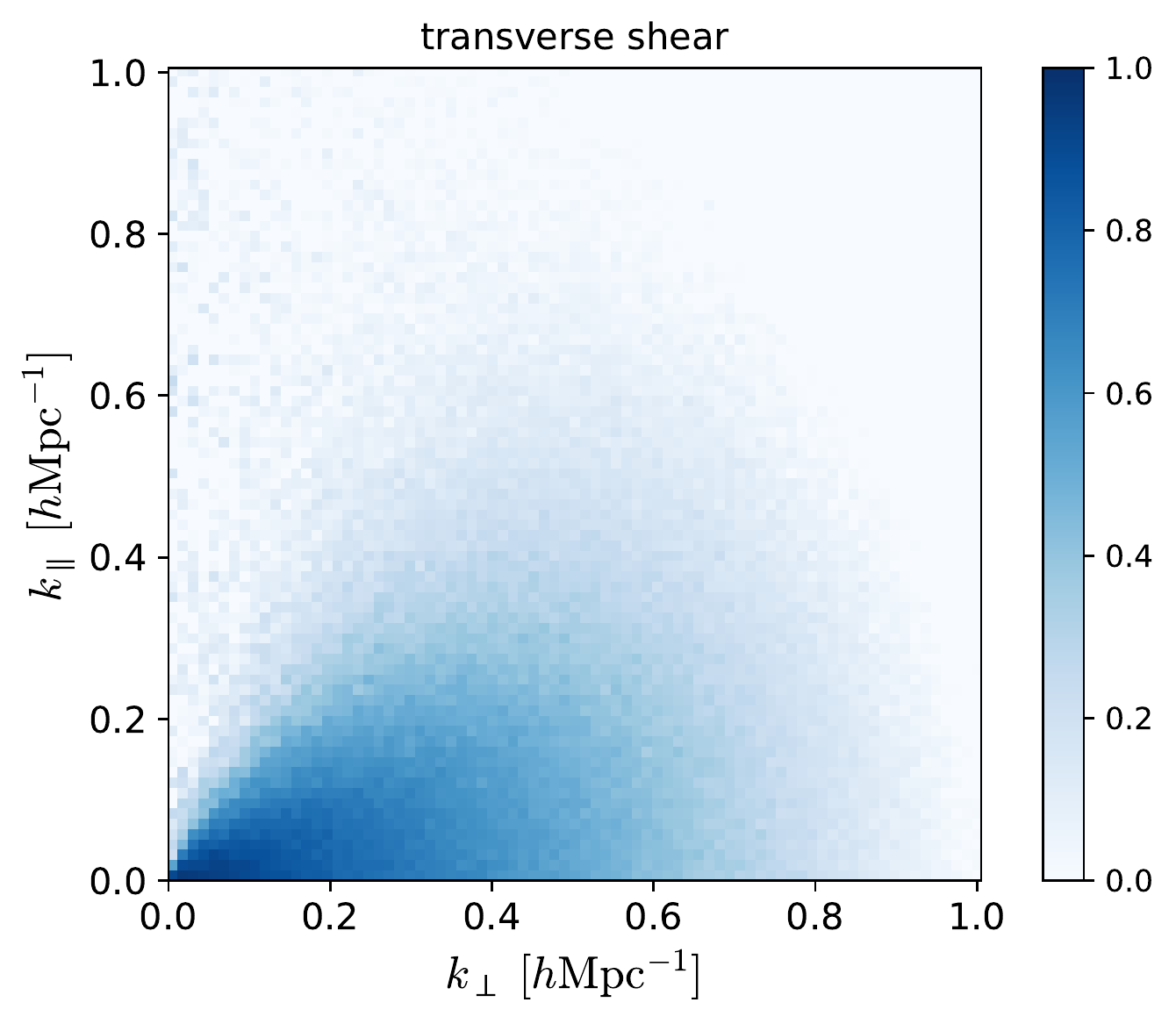}
\includegraphics[width=0.47\textwidth]{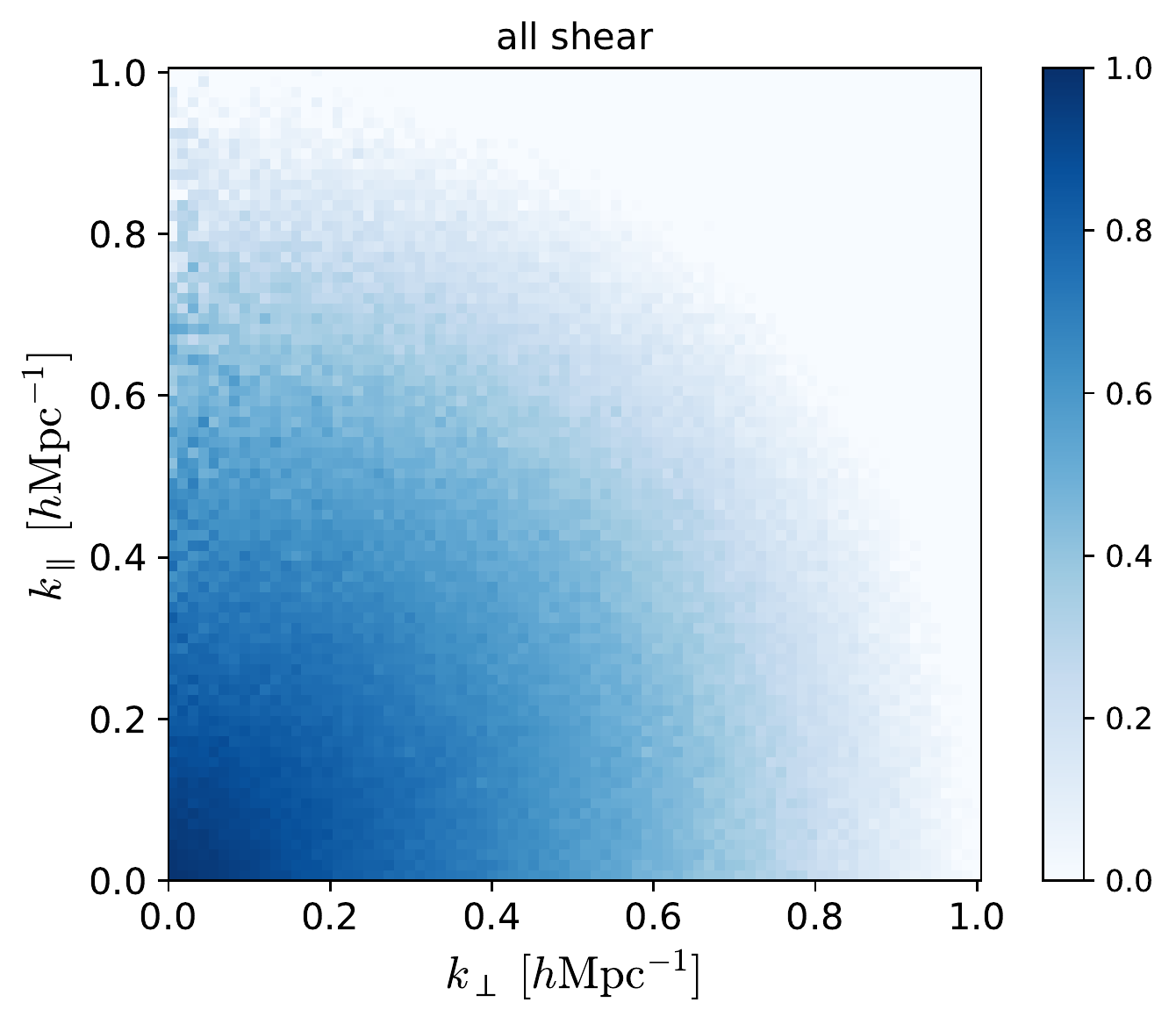}
\caption{
The two-dimensional cross-correlation coefficient $r(k)$ of the reconstructed field $\delta_r$ and the dark matter density field $\delta$ for number density $\bar{n}=4.9\times10^{-2}\;(\mathrm{Mpc}/h)^{-3}$, using all shear fields (Top) and transverse shear (Bottom).
\label{fig:xcc_2d}}
\end{figure}

We plot the two-dimensional cross correlation coefficient for reconstruction with $\bar{n}=4.9\times10^{-2}\;(\mathrm{Mpc}/h)^{-3}$ in Fig.~\ref{fig:xcc_2d}.
For comparison, we also plot the correlation coefficient for reconstruction with only the transverse shears as former methods \citep[][]{2012Tides,2016Tides,2019MNRAS.486.3864K}.
Using only two transverse shear fields leads an anisotropic correlation in Fourier space.
This is due to that transverse shear only probes the transverse modes directly, while the radial modes are inferred by the change of the transverse shear along the line-of-sight direction \citep[see][for more details about this]{2016Tides}.
Including all shear field improves the performance and reduces the large noise in the high $k_\parallel$ regime.

\begin{figure}[tb!]
\includegraphics[width=0.47\textwidth]{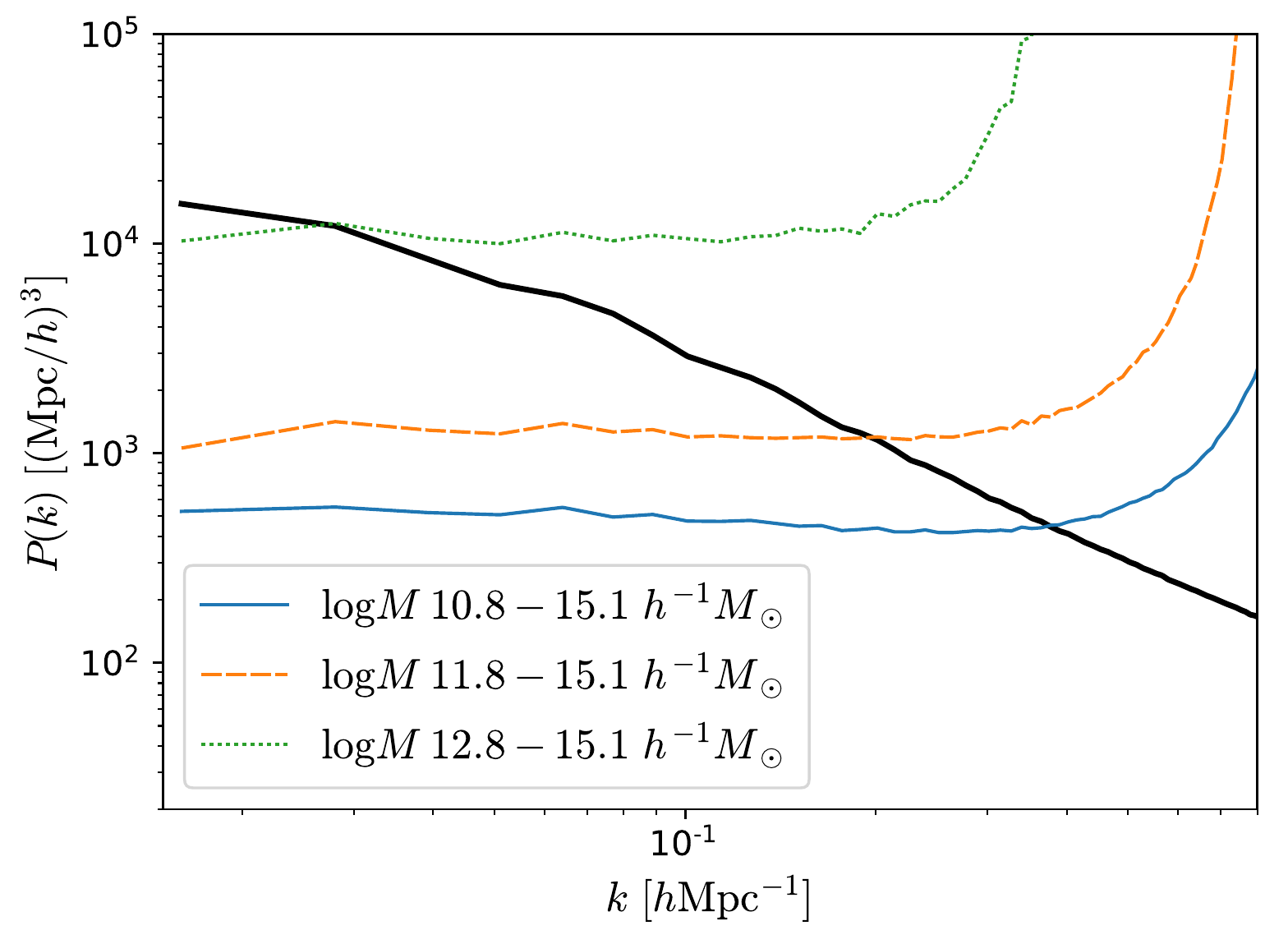}
\caption{The noise power spectrum for reconstruction with halo number density $\bar{n}=4.9\times10^{-2}$, $6.3\times10^{-3}$, and $5.9\times10^{-4}\ (\mathrm{Mpc}/h)^{-3}$. 
The reconstruction noise power level is approximately scale independent on large scales.
\label{fig:noise}}
\end{figure}

In the data analysis, we also need to model the power spectrum of stochastic noise.
The reconstruction noise is defined by 
\begin{equation}
    N(\bmk)=\delta_r(\bmk)/C(\bmk)-\delta(\bmk),
\end{equation}
where $\delta_r$ is the reconstructed field and $\delta$ is the dark matter density field and the noise power spectrum is then given by
\begin{equation}
    P_N(k)=P_{\delta_r}(k)/C^2(k)-P_\delta(k).
\end{equation}
In Fig.~\ref{fig:noise}, we plot the noise power spectrum for different cases.
We see that the reconstruction noise power level is nearly scale-independent at $k<0.1
\;h/\Mpc^{-1}$ and diverges on small scales.
This can be interpreted as arising from the decreasing of reconstruction bias $C(k)$ on small scales.
Similarly, we expect that for modes with $k<0.1\;h/\Mpc$, the noise power spectrum is constant to a good approximation, which can be modeled as a constant term in the cosmological data analysis. 
The noise level is generally a few times larger than the shot noise of the halo catalogs used for reconstruction. 
However, this field provides an independent tracer of the large-scale density which can be used to cancel cosmic variance in the galaxy density field.

\section{Discussion and conclusion}
\label{sec:dissconcl}

In this paper, we present a new tidal reconstruction algorithm using all five shear fields in the three-dimensional space, which improves upon the previous method using only the transverse shear fields. 
We have applied the new algorithm to dark matter halos from simulations.
The correlation coefficient with the dark matter density is larger than 0.9 at $k\lesssim 0.1\;h\Mpc^{-1}$ for the two dense samples and for the sparse sample the correlation coefficient is still larger than 0.5 at $k\lesssim0.1\;h\Mpc^{-1}$.
If we already have the full dark matter density field from observations, there is no need to do tides.
We can just take the original field.
However, in 21~cm intensity mapping surveys, some of the small $k$ modes can not be measured, so tides are clearly useful and recover the lost radial information.

Such high correlation with the dark matter density makes the reconstructed field an ideal field for the multi-tracer method \citep[][]{2009MT} (see also \citet{2020arXiv200708472D}).
Although for the BOSS \citep[][]{2017MNRAS.470.2617A} and eBOSS \citep[][]{2021PhRvD.103h3533A} samples, the tidal reconstruction is not efficient because of the high shot noise, the near term and future surveys, including DESI \citep[][]{2016arXiv161100036D}, Euclid \citep[][]{2018LRR....21....2A,2020A&A...642A.191E}, SPHEREx \citep[][]{2014arXiv1412.4872D}, MegaMapper \citep[][]{2019BAAS...51g.229S,2019BAAS...51c..72F}, and Billion Object Apparatus \citep[][]{2016arXiv160407626D}, etc will have a significantly higher number density. 
The SDSS Main Galaxy Sample has a much higher number density than the BOSS and eBOSS samples \citep[][]{2005AJ....129.2562B}, which is ideal for performing tidal reconstruction and testing the algorithm.
We plan to explore this in a future work.

The tidal reconstruction acquires a large-scale linear bias, because the real tidal coupling strength can not be predicted for density fluctuations in the nonlinear regime.
However, on large scales $k<0.1\;h\Mpc^{-1}$ that are important for $f_{\mathrm{NL}}$ constraints, the reconstruction bias is constant to an excellent approximation.
If only using the modes in the quasi-linear regime where perturbative description is still valid, the reconstruction bias can in principle be predicted using perturbation theory \citep[e.g.][]{2018JCAP...07..046F,2020PhRvD.101h3510L,2020arXiv200700226L,2020arXiv200708472D}, but at the cost of losing a lot of small-scale information and increased reconstruction noise. 
Since we are trying to reconstruct the large-scale mode through its gravitational effects on the small scale matter fluctuations and the performance is dominated by the numerous modes in the squeezed limit, we expect the tidal reconstruction is not directly $f_{\mathrm{NL}}$-sensitive.
However, more detailed studies using very large volume simulations with primordial non-Gaussianity will be needed.
We plan to study this in future.

The cross spectrum of the reconstructed field with the original field is a bispectrum of the nonlinear density, while the power spectrum of the reconstructed field is a trispectrum.
The tidal reconstruction method is indeed using the nonlinear statistics to improve the measurement of large-scale structures.

The reason we originally only used transverse modes is because redshift space distortions will affect the other components \citep[][]{2012Tides,2016Tides}.
Using all tidal shear fields clearly improves the reconstruction performance, especially in the high $k_\parallel$ regime. 
However, the three shear fields involving radial direction may be affected by the redshift space distortions.
We plan to investigate this effect in a forthcoming paper.

In this paper we study the performance with the three-dimensional fields.
This method can also be applied to photometric galaxies, with a number of future photometric surveys including Vera Rubin Observatory/LSST \citep[][]{2009arXiv0912.0201L}, Euclid \citep[][]{2018LRR....21....2A,2020A&A...642A.191E}, and Roman Space Telescope/WFIRST \citep[][]{2019arXiv190205569A}.
In that case, most information would be coming from the two transverse shears due to the large photo-$z$ errors.
However, the benefits may be even greater for photometric surveys, where no spectroscopy is needed and hence can have a higher number density with a significantly small investments.
The effect of photometric redshift errors and redshift space distortions on tidal reconstruction will be explored in a future paper.

\begin{acknowledgments}
We thank Marcel Schmittfull for enlightening discussions and sharing the simulations.
We receive support from Natural Sciences and Engineering Research Council of Canada (NSERC) [funding reference number RGPIN-2019-067, 523638-201, CITA 490888-16], Canadian Institute for Advanced Research (CIFAR), Canadian Foundation for Innovation (CFI), Simons Foundation, and Alexander von Humboldt Foundation. 
The Dunlap Institute is funded through an endowment established by the David 
Dunlap family and the University of Toronto.
Research at the Perimeter Institute is supported by the Government of Canada 
through Industry Canada and by the Province of Ontario through the Ministry of 
Research and Innovation.
\end{acknowledgments}

%





\bibliography{ms}{}
\bibliographystyle{aasjournal}



\end{document}